\documentclass[jcp,aps,twocolumn,eqsecnum,a4paper,ltxgrid]{revtex4-1}

\usepackage{graphicx}
\usepackage[sumlimits]{amsmath}
\usepackage{amsfonts}
\usepackage{amssymb}
\usepackage{dcolumn}
\usepackage{subdepth}
\usepackage{color}
\usepackage{enumitem}

\usepackage{soul}
\setstcolor{red}

\newcommand{\mOp}[1]{\hat{#1}}
\newcommand{\mVec}[1]{\boldsymbol{\mathrm{#1}}}
\newcommand{\mBra}[1]{\langle #1 |}
\newcommand{\mKet}[1]{| #1 \rangle}

\newcommand{\mAve}[1]{\left \langle #1 \right \rangle}
\newcommand{\Eq}[1]{Eq.\,(\ref{#1})}
\newcommand{\Eqs}[1]{Eqs.\,({#1})}
\newcommand{\Sec}[1]{Sec.\,\ref{#1}}
\newcommand{\mMat}[1]{{\boldsymbol{\mathrm{#1}}}}

\newcommand{\FT}[1]{\check{#1}}
\newcommand{\e}{\mathrm{e}}
\renewcommand{\i}{\mathrm{i}}
\newcommand{\diff}{\mathrm{d}}
\newcommand{\Fig}[1]{Fig.\,\ref{#1}}

\newcommand{\cm}{cm$^{-1}$}
\renewcommand{\Re}{\mathrm{Re}}

\begin{document}
	\title{A time-correlation function approach to nuclear dynamical effects in X-ray spectroscopy}
	
	\author{Sven Karsten}
	\address{Institute of Physics, University of Rostock, Albert-Einstein-Str. 23-24, 18059 Rostock, Germany}
	\author{Sergey I.\ Bokarev}
		\email{sergey.bokarev@uni-rostock.de}
	\address{Institute of Physics, University of Rostock, Albert-Einstein-Str. 23-24, 18059 Rostock, Germany}
	\author{Saadullah G. Aziz}
	\address{Chemistry Department, Faculty of Science, King Abdulaziz University, 21589
		Jeddah, Saudi Arabia}
	\author{Sergei D. Ivanov}
		\email{sergei.ivanov@uni-rostock.de}
	\address{Institute of Physics, University of Rostock, Albert-Einstein-Str. 23-24, 18059 Rostock, Germany}
	\author{Oliver K\"uhn }
	\address{Institute of Physics, University of Rostock, Albert-Einstein-Str. 23-24, 18059 Rostock, Germany}
	

\begin{abstract}
Modern X-ray spectroscopy has proven itself as a robust tool for probing the electronic structure of atoms in complex environments.
Despite working on energy scales that are much larger than those corresponding to nuclear motions, taking nuclear dynamics and the associated nuclear correlations into account may be of importance for X-ray spectroscopy.
Recently, we have developed an efficient protocol to account for nuclear dynamics in X-ray absorption and resonant inelastic X-ray scattering spectra~[Karsten~\textit{et~al.} arXiv:1608.03436], based on ground state molecular dynamics accompanied with state-of-the-art calculations of electronic excitation energies and transition dipoles.
Here, we present an alternative derivation of the formalism and elaborate on the developed simulation protocol on the examples of gas phase and bulk water.
The specific spectroscopic features stemming from the nuclear motions are analyzed and traced down to the dynamics of electronic energy gaps and transition dipole correlation functions.
The observed tendencies are explained on the basis of a simple harmonic model and the involved approximations are discussed.
The method represents a step forward over the conventional approaches treating the system in full complexity and provides a reasonable starting point for further improvements.
\end{abstract}

\maketitle
\section{Introduction}
Understanding complex phenomena arising in physical chemistry requires unraveling the underlying processes on an atomistic level.
Due to the energetic separation of the core levels of different elements and the compact nature of the corresponding orbitals, X-ray spectroscopy can reveal highly local and element-specific information on the electronic structure of an absorbing atom and on its interaction with the environment.~\cite{Stohr-Book-1992}
In particular, X-ray absorption spectra (XAS) probe those electronic transitions, where a core electron is excited to the unoccupied molecular orbitals (MOs), whereas resonant inelastic X-ray scattering (RIXS)  detects the emission signal resulting from the refill of the core hole by electrons from occupied MOs.
Thus, the combination of both approaches provides detailed information on the electronic structure of a sample in different 
aggregation states~\cite{degroot2008,Winter2006} leading to the popularity of the two methods.

Along with the improvement of the instrumental resolution in X-ray spectroscopic studies, not only electronic but also the vibrational transitions have recently attracted growing attention.~\cite{Rubensson2013, Guillemin2013} In Ref.~\citenum{Dong2013}, RIXS has been even regarded to be a complementary technique to the conventional vibrational infra-red and Raman spectroscopies.
Although X-ray scattering processes occur during few femtoseconds due to the typically short lifetimes of core-excited states, indications of ultrafast nuclear dynamics could be observed.
For instance, 
the inclusion of vibronic effects in RIXS spectra as well as coherences between vibrational levels has  been found essential for the correct assignment of spectral features in simple model systems.~\cite{Hennies2005, Ljungberg2011}
Further, dissociative dynamics on the timescale of the RIXS scattering process~\cite{Bohinc2013,Pietzsch2011,Hennies2010} has also been detected.

Remarkably, the RIXS spectra of liquid water and alcohols initiated active debates in the last decade due to the splitting of the 1b$_1$ band, which is not observed in the photoelectron spectra.~\cite{Lange2013, Schreck2014,Fransson-CR-2016}
This peculiar effect received controversial interpretation resulting in six hypotheses,~\cite{Sellberg2015} with four of them involving different aspects of nuclear dynamics, such as ultrafast dissociation and H-bond dynamics.
Thus, the necessity for a robust theoretical treatment of nuclear vibrational effects becomes apparent.

Since solving the electronic-nuclear Schr\"odin\-ger equation is feasible only for rather small model systems,~\cite{paramonov12_11388} different  approximate schemes are commonly applied.
Frequently, electronic spectra are obtained via single point electronic structure calculations combined with phenomenological broadening for the vibrational environmental effects, thereby neglecting peculiarities of the underlying microscopic dynamics, see, e.g., Ref.~\citenum{Leetmaa2010}.
A popular extension to this approach that explicitly includes nuclear vibrations is to assume the shifted harmonic potentials for the initial and final electronic states, leading to the analytical Franck-Condon description.~\cite{Kuehn-Book, Wachtler2012, Sakko2011, Hennies2005}
However, this approach is not appropriate for cases where strong anharmonicities, bond formation or cleavage, and/or pronounced conformational changes are present.
Here, real-time propagation of a nuclear wavepacket on pre-calculated potential energy surfaces improves on the purely harmonic description.~\cite{Saek2003, Ljungberg2011, Wachtler2012, Couto-PRA-2016}
Nevertheless, the construction of such
multidimensional potentials as well as the wavepacket
propagation itself are practically unfeasible for large numbers of highly-excited electronic states (relevant for RIXS spectra) and nuclear degrees of freedom.

Classical molecular dynamics (MD) simulations employing forces according to the electronic ground state have proven themselves as a versatile approach to incorporate and analyze the spectral fingerprints of nuclear dynamics in infrared to ultraviolet  spectroscopic ranges.~\cite{Kuehn-Book, Marx-Book-2009, Ivanov-PCCP-2013, zhuang09_3750}
Extending this methodology to the X-ray range and performing quantum-chemical static point calculations for each MD snapshot allows one to sample nuclear distributions in the phase space, leading to a more realistic description of conformational and environmental effects.~\cite{Sun2011, Jena2015, Weinhardt2015, Leetmaa2010}
However, the sampling approach is capable of describing a distribution of structural motifs only, without any (time-ordered) nuclear correlation and a truly time-domain approach is needed to enable both statical and dynamical effects.
Interestingly,
these effects could be disentangled from each other experimentally, employing RIXS with 
excitation pulses strongly detuned from the resonance.~\cite{Schreck2014}

Recently, we have suggested a time-domain approach for calculating XAS and RIXS amplitudes from the time evolution provided by electronic ground state MD simulations and have demonstrated that  nuclear correlation effects are indeed essential on the example of a gas phase water molecule.~\cite{Karsten-arXiv-2016} 
Here, we present an alternative but still rigorous derivation for the formalism, starting from  Fermi's Golden rule and the Kramers-Heisenberg expression for XAS and RIXS amplitudes, respectively, followed by employing the interaction representation picture and the dynamical classical limit~\cite{Kuehn-Book,Mukamel-Book} as is described in \Sec{sec:Theory}.  
To retain nuclear correlations, one has to trace the entire manifold of relevant electronic levels along the MD trajectory
in order to eliminate possible order and phase alterations.
A fully-automated procedure serving this purpose has been developed
as is illustrated in Sections~\ref{sec:Methodology} and \ref{sec:computational_details}.
The established protocol is exemplified by oxygen K-edge XAS and RIXS of gas phase and bulk liquid water.
However, in the present article we mainly focus on methodological aspects rather than on the particular application and, thus, the obtained results have been compared against those of the sampling approach, not focusing on the peculiarities of water X-ray spectra, see \Sec{sec:results}.
The developed correlation-function technique should provide a step forward in the description of nuclear dynamical effects in various X-ray spectroscopies.
It can be applied to generally anharmonic systems treating them in full dimensionality, that is without scanning potential energy surfaces.
Respective conclusions as well as limitations of the method are summarized in \Sec{sec:limitations} and \Sec{sec:conclusions}, respectively.
\section{Theory}
\label{sec:Theory}
\subsection{Setting the stage}
\label{sec:Setting_stage}
Let us consider a molecular system consisting of electrons represented by Cartesian
positions $\mOp{\mVec{r}}$ and momenta $\mOp{\mVec{p}}$ and nuclei described, respectively, by $\mOp{\mVec{R}}$ and $\mOp{\mVec{P}}$.
In the framework of the Born-Oppenheimer approximation (BOA), an eigenstate of the total Hamiltonian, having the energy $\epsilon_\alpha$, factorizes as $\mKet{\alpha}=\mKet{a}\mKet{A}$; here and in the following nuclear states denoted with a capital letter correspond to an electronic state indicated with the same small letter, e.g.\ a set of nuclear states $\mKet{A}$ belongs to the electronic state $\mKet{a}$.
The respective electronic energies are given as the solutions of the electronic time-independent Schr\"odinger equation (TISE)
\begin{equation}
 \label{eq:BOA2}
 \mOp{H}_\mathrm{el}(\mOp{\mVec{r}}, \mOp{\mVec{p}}, \mOp{\mVec{R}})\mKet{a}=\mOp{E}_{a}(\mOp{\mVec{R}})\mKet{a}
 \enspace,
\end{equation}
where $\mOp{H}_\mathrm{el}$ is the electronic Hamiltonian, see, e.g., Ref.~\citenum{Kuehn-Book}.
Finally, the nuclear state $\mKet{A}$ is an eigenstate of the Hamiltonian $\mOp{H}_{a}(\mOp{\mVec{R}},\mOp{\mVec{P}}) := \mOp{H}_{\mathrm{nuc}}(\mOp{\mVec{R}},\mOp{\mVec{P}})+\mOp{E}_{a}(\mOp{\mVec{R}})$
with the eigenvalue $\epsilon_{A}$ 
\begin{equation}
\label{eq:nuclear_eigenstate_relation}
\mOp{H}_{a}\mKet{A} =\epsilon_{A}\mKet{A}
\enspace,
\end{equation}
with $\mOp{H}_{\mathrm{nuc}}$ consisting of the nuclear kinetic and potential (Coulomb) energy operators~\cite{Kuehn-Book}.

In the following, atomic units will be used and the arguments of the operators will be skipped unless the dependency has to be emphasized.
In general, the indices stemming from electronic bra-states appear in superscript, whereas the subscript ones correspond to the ket-states.
For the sake of brevity, the notation $c^{\xi} := c_{\xi}^{*}$ for any indexed complex quantity and 
$\mOp{A}^{\xi} := \mOp{A}_{\xi}^{\dagger}$ for any indexed operator (vector) is used throughout the manuscript.
Furthermore, the Einstein notation is employed, that is, indices that appear as a subscript {\it and} as a superscript are summed over.
%

\subsection{The XAS amplitude}
\label{sec:XAS_amplitude}

In this section, the derivation of a time-domain expression for the XAS amplitude is presented in detail.
We opted to present it because of its relative simplicity, rather than the derivation of a perhaps more interesting but cumbersome expression for the RIXS amplitude.
Still the present derivation contains all the necessary steps and can thus serve as a roadmap for deriving the RIXS amplitude in the time domain.
The latter is presented here only schematically with all the details left for the Supplement.
In contrast to the derivation presented in the Supplement of Ref.~\citenum{Karsten-arXiv-2016}, we do not employ optical response functions here.

\subsubsection{The quantum expression}

\begin{figure}
\includegraphics[scale=0.4]{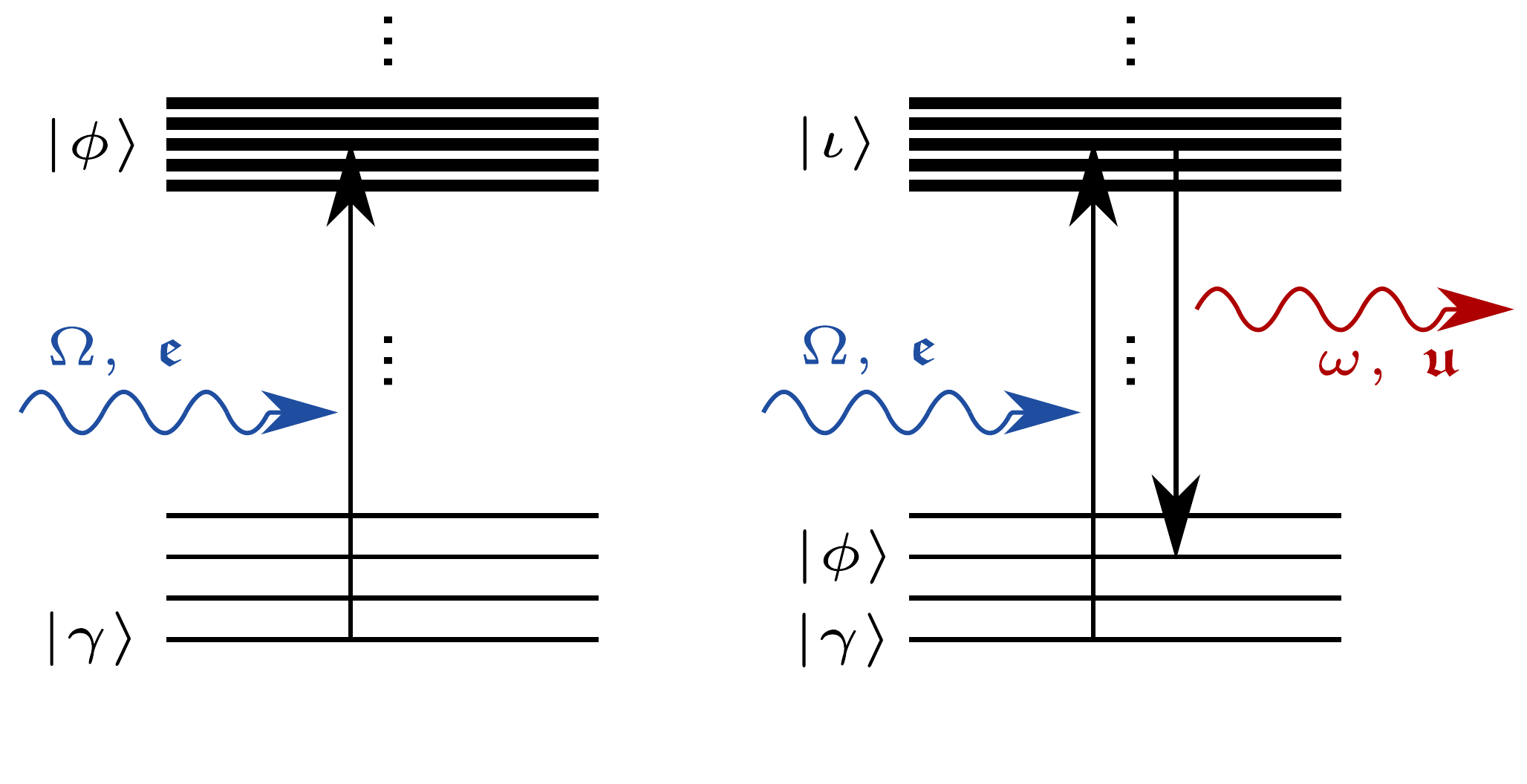}
\caption{\label{fig:sketch}
Schematic sketch of XAS (left) and RIXS (right) processes, see text.
}
\end{figure}

The process under study consists of exciting the system from an initial state $\mKet{\gamma}$ to a final state $\mKet{\phi}$ by absorbing light with angular frequency $\Omega$ and polarization vector with the Cartesian components $\mathfrak{e}_{\xi}$, see left panel in \Fig{fig:sketch}. 
Following   Fermi's Golden rule, the X-ray absorption spectrum is proportional to the XAS amplitude that can be written down with the help of the notations introduced in the previous section as
\begin{equation}
\label{eq:XAS}
\mathcal{X}(\Omega) =
 \sum_{\gamma,\phi} \frac{\e^{-\beta \epsilon_{\gamma}}}{Z}\left | \mBra{\phi}  \mathfrak{e}^{\xi} \mOp{d}_\xi \mKet{\gamma}  \right|^{2} \delta \left( \Omega-(\epsilon_{\phi}-\epsilon_{\gamma}) \right)
\enspace, 
\end{equation}
where $\beta:=1/(k_{\mathrm{B}}T)$ is the inverse temperature, $k_{\mathrm{B}}$ is the Boltzmann constant, $Z:=\sum_{\gamma}\exp \left[ -\beta \epsilon_{\gamma} \right]$ is the canonical partition function, and $\mOp{d}_{\xi}$ is the $\xi$-th component of the dipole operator. 
The Dirac $\delta$-function ensures the energy conservation during the process. 

In order to formulate the time-domain version of \Eq{eq:XAS}, the $\delta$-function is represented as the time integral
\begin{equation}
\label{eq:delta_time_domain}
\delta\left(\Omega-\Omega_{0}\right)=\frac{1}{2\pi}\intop_{-\infty}^{\infty}\diff t\,\e^{\i\left(\Omega-\Omega_{0}\right)t}
\enspace ,
\end{equation}
yielding
\begin{align}
 \label{eq:XAS_last_before_BOA}
 \mathcal{X}(\Omega)=  &\frac{1}{2\pi Z} \sum_{\gamma,\phi} \intop_{-\infty}^{\infty}\diff t\,
     \e^{\i\left(\Omega-(\epsilon_{\phi}-\epsilon_{\gamma})\right)t} \nonumber \\
 \times & \e^{-\beta \epsilon_{\gamma}} \mBra {\gamma} \mOp{d}^{\eta}  \mathfrak{e}_{\eta}
    \mKet{\phi} \mBra{\phi}  \mathfrak{e}^{\xi} \mOp{d}_\xi \mKet{\gamma} 
\end{align}
in the fashion similar to the derivation of the infrared and UV/Vis spectroscopies~\cite{Heller1978,lawrence2002,harder2005,Kuehn-Book,McQuarrie-Book}.
Alternatively, this expression follows from time-dependent perturbation theory as is exercised in Ref.~\citenum{Mukamel-Book}, see also Ref.~\citenum{Karsten-arXiv-2016}.
Note that in order to keep notations as compact as possible the transition dipoles are assumed to be projected on the polarization vector and the corresponding coordinate indices $\xi$ and $\eta$ are dropped throughout the theory section without any loss of generality. They will be restored in the final expressions.

Applying the BOA leads to the aforementioned factorization of the states: $\mKet{\gamma}=\mKet{g}\mKet{G}$,
and $\mKet{\phi}=\mKet{f}\mKet{F}$, which after rearranging the complex exponential in \Eq{eq:XAS_last_before_BOA} leads to
\begin{align}
 \label{eq:XAS_after_BOA}
\mathcal{X}(\Omega)&=  \frac{1}{2\pi Z} \sum_{G,F}  \intop_{-\infty}^{\infty}\diff t\,\e^{\i\Omega t}  \nonumber \\
  &\times\mBra{G}  \e^{-\beta \epsilon_{G}} \e^{\i\epsilon_{G}t}\mOp{D}^{g}_f   \e^{-\i\epsilon_{F}t} \mKet{F} \mBra{F} \mOp{D}^{f}_{g} \mKet{G}
\enspace,
\end{align}
where $\mOp{D}_{b}^{a}(\mOp{\mVec{R}}) :=\mBra{a}\mOp{d}(\mOp{\mVec{r}},\mOp{\mVec{R}})\mKet{b}$
is the electronic transition dipole moment that remains an operator in the nuclear space.
In order to obtain the time evolution for the dipole operator, one may 
substitute the energies $\epsilon_{F}$ and $\epsilon_{G}$, being the eigenvalues of $\mOp{H}_{f}$ and
$\mOp{H}_{g}$, correspondingly, see \Eq{eq:nuclear_eigenstate_relation}, by the operators themselves.
Since after this substitution there is no dependence on the final nuclear states left in \Eq{eq:XAS_after_BOA}, one may 
employ the closure for $\mKet{F}$ to eliminate the projectors $\mKet{F} \mBra{F}$, leading to
\begin{align}
\label{eq:XAS_before_scattering}
\mathcal{X}(&\Omega)=  \frac{1}{2\pi Z}   \intop_{-\infty}^{\infty}\diff t\,\e^{\i\Omega t}  \nonumber \\
& \times \sum_{G} \mBra {G} \e^{-\beta \mOp{H}_g} \e^{\i\mOp{H}_{g}t}\mOp{D}^{g}_f   \e^{-\i\mOp{H}_{f}t}
   \mOp{D}^{f}_{g} \mKet{G}   \enspace.
\end{align}
The mission to obtain a practical time-domain analogue of \Eq{eq:XAS} would be accomplished, if $\mOp{H}_{f}$ coincided with $\mOp{H}_{g}$. This difference can be circumvented by the following steps.
First, one rewrites
$\mOp{H}_{f}=\mOp{H}_{0}+\Delta \mOp{E}_{f0}$, $\Delta \mOp{E}_{f0}:=\mOp{E}_{f}-\mOp{E}_{0}$ with the index $0$ standing for the electronic ground state. 
Second, the energy gap is treated as a perturbation operator that enables switching to the interaction representation~\cite{Kuehn-Book,Mukamel-Book}
\begin{equation}
\label{eq:interaction_representation}
\e^{-\i\hat{H}_{f}t}=\e^{-\i\hat{H}_{0}t}\hat{S}_{f}(t,0)
\enspace,
\end{equation}
where the scattering operator is defined as
\begin{equation}
\label{eq:scattering_op_definition}
 \hat{S}_{f}(t,t_{0}):=\exp_{+}\left[ -\i\intop_{t_{0}}^{t}\diff\tau \, \Delta \mOp{E}_{f0}(\tau)  \right]
\enspace.
\end{equation}
The symbol $\exp_{+}$ in \Eq{eq:scattering_op_definition} represents a (positively) time-ordered exponential. 
The time argument of the integrand therein stands for the time evolution according to the Heisenberg equation of motion with respect to the electronic ground state Hamiltonian, $\hat{H}_0$. 
The equations for $\mKet{g}$ can be obtained by taking the adjoint from \Eqs{\ref{eq:interaction_representation},\ref{eq:scattering_op_definition}} and substituting $f$ by $g$.
Importantly, the choice of the electronic ground state as the reference is 
motivated by the  initial condition before absorbing a photon and is 
beneficial in view of potential use of classical MD methods, which are available mostly for the ground state.

Utilizing the interaction representation, \Eq{eq:interaction_representation}, for the XAS amplitude yields
\begin{align}
\mathcal{X}(&\Omega)=  \frac{1}{2\pi Z}  \intop_{-\infty}^{\infty}\diff t\,\e^{\i\Omega t} \sum_{G}  \nonumber \\
 & \times \mBra {G} \e^{-\beta \mOp{H}_g(0)} \hat{S}^{g}(t,0)\mOp{D}^{g}_f(t)\hat{S}_{f}(t,0)\mOp{D}^{f}_{g}(0) \mKet{G}
 \enspace.
\end{align}
Introducing the ``dressed'' dipole moment 
\begin{equation}
\mOp{M}^{a}_{b}(t,0):=\hat{S}^{a}(t,0)\mOp{D}^{a}_{b}(t)\hat{S}_{b}(t,0)
\end{equation}
leads to a compact form for the XAS amplitude
\begin{align}
\label{eq:XAS_QM}
\mathcal{X}(&\Omega)=  \frac{1}{2\pi Z}  \intop_{-\infty}^{\infty}\diff t\,\e^{\i\Omega t}    \nonumber \\
& \times  \sum_{G}  \mBra {G} \e^{-\beta \mOp{H}_g(0)} \mOp{M}^{g}_f(t,0) \mOp{M}^{f}_{g}(0,0) \mKet{G}   
\enspace,
\end{align}
where we have used that $\mOp{M}^{f}_{g}(0,0)\equiv\mOp{D}^{f}_{g}(0)$. 
%

\subsubsection{The dynamical classical limit}
\label{sec:DCL}
%
In order to utilize well-established classical MD methods in combination with  state-of-the-art quantum chemistry tools, the quantum mechanical expression for the XAS amplitude, \Eq{eq:XAS_QM}, is subjected to the dynamical classical limit for the nuclei~\cite{Mukamel-Book,Kuehn-Book}.
In particular, the nuclei are represented by point particles and the operators are replaced by continuous dynamical functions.
Consequently, the trace over the initial nuclear states $\mKet{G} $ is substituted by an integral over the phase space
and time-evolved operators become dynamic functions.
Naturally, the time-ordering of exponentials in \Eq{eq:scattering_op_definition} becomes irrelevant and disappears.

Performing all these replacements in Eq.\,(\ref{eq:XAS_QM}) results in
\begin{align}
\mathcal{X}(&\Omega)=  \frac{1}{2\pi Z}  \intop_{-\infty}^{\infty}\diff t\,\e^{\i\Omega t}    \nonumber \\
&  \times  \iint \diff\mVec{R}_0 \diff\mVec{P}_0  \e^{-\beta {H}_g(0)} {M}^{g}_f(t,0) {M}^{f}_{g}(0,0)
\enspace,
\end{align}
where $Z$ becomes the sum of \textit{classical} partition functions corresponding to potential energy surfaces of all electronic states in question,
$Z=\sum_{g}\iint \diff\mVec{R}_0 \diff\mVec{P}_0 \, \exp[-\beta H_{g}(0)]$.
Since the aim is to use classical MD methods, the observables should have the form of a canonical ensemble average,
$\langle \bullet \rangle$,
with respect to the Hamilton function of the electronic ground state
%
\begin{equation}
\left\langle A\right\rangle  :=\frac{1}{Z_0}\iint \diff\mVec{R}_0\diff\mVec{P}_0 \, 
    \e^{-\beta H_{0}(\mVec{R}_0,\mVec{P}_0)} \thinspace A(\mVec{R}_0,\mVec{P}_0)
\enspace ,
\end{equation}
where $Z_{0}:=\iint \diff\mVec{R}_0\diff\mVec{P}_0 \thinspace\exp[-\beta H_{0}(\mVec{R}_0,\mVec{P}_0)]$.
Using the standard trick of adding and subtracting $H_0$ and performing some simple algebra leads to
%
\begin{equation}
\label{eq:XAS_final}
\mathcal{X}(\Omega)= \frac{1}{2\pi}  \intop_{-\infty}^{\infty}\diff t\,\e^{\i\Omega t} 
  \mAve{ \mathcal{W}_g(0) {M}^{g \eta}_f(t,0) \mathfrak{e}_{\eta}  \mathfrak{e}^{\xi}{M}^{f}_{\xi g}(0,0) }
\enspace,
\end{equation}
where the omitted polarization vectors have been restored.
Here, the weighting function is
\begin{equation}
\label{eq:weighting_factor}
 \mathcal{W}_g(0):=\e^{-\beta[\Delta E_{g0}(0)]}/ \mathcal{Z}
\end{equation}
with the normalization factor
\begin{equation}
\label{eq:norm_factor}
  \mathcal{Z}:=\mAve{ \sum_{g}\e^{-\beta[\Delta E_{g0}(0)]}} 
\enspace ,
\end{equation}
see Supplement.
Equation~(\ref{eq:XAS_final}) for the XAS amplitude has the desired form
of the Fourier transform of an equilibrium time correlation function.

\subsection{The RIXS amplitude}
\label{sec:RIXS_DCL}

It was pointed out in the Introduction, that the XAS spectra considered above correspond to one-photon processes,
whereas more information can be obtained from a two-photon process, e.g.\ RIXS.
Here, the system is first excited from the initial state $\mKet{\gamma}$ to an intermediate state $\mKet{\iota}$
by absorbing light with a frequency $\Omega$ and a polarization vector with the components $\mathfrak{e}_{\xi}$,
see right panel in \Fig{fig:sketch}.
Second, the system transits from the state $\mKet{\iota}$ to the final state $\mKet{\phi}$
by emitting light with the frequency $\omega$ and the polarization vector with components $\mathfrak{u}_{\eta}$.

Following Kramers and Heisenberg, the RIXS amplitude reads~\cite{kramers-heisenberg}
\begin{align}
\label{eq:Kramers_Heisenberg}
\mathcal{R}(\Omega,\omega)=\sum_{\gamma,\phi} \frac{\e^{-\beta \epsilon_{\gamma}}}{Z}
\left| \sum_{\iota}\frac{\mBra{\phi} \mOp{d}^{\eta} \mathfrak{u}_{\eta} \mKet{\iota}\mBra{\iota}\mathfrak{e}^{\xi} \mOp{d}_{\xi}\mKet{\gamma}}{\Omega-(\epsilon_{\iota}-\epsilon_{\gamma})+\i\Gamma_{\iota}}\right|^{2} \nonumber \\
\times\delta\left(\Omega-(\epsilon_{\phi}-\epsilon_{\gamma}+\omega)\right)
\enspace,
\end{align}
where the dephasing rate $\Gamma_{\iota}$ takes into account the finite lifetime of the state $\mKet{\iota}$.
This implies that intricate electron-nuclear dynamics therein, including non-radiative relaxation mechanisms, such as the Auger one, is modeled by a simple mono-exponential decay.
Note that the sum in \Eq{eq:Kramers_Heisenberg} is under the square, thereby coherences are taken into account.

In order to get a classical time-domain expression for RIXS, the same pathway as for XAS is followed,
as is summarized below and detailed in the Supplement. 
To start, the integral representation of a $\delta$-function, \Eq{eq:delta_time_domain}, is employed.
Additionally, the denominator in \Eq{eq:Kramers_Heisenberg} is recast into the time domain via
\begin{equation}
\label{eq:damping_time_domain}
\frac{1}{\omega-\omega_{0}\pm\i\Gamma_\alpha}=\mp\i\intop_{-\infty}^{\infty}\diff t\,\e^{\pm\i\left(\omega-\omega_{0}\right)t}\Delta_\alpha(t)\enspace,
\end{equation}
with the damping function 
\begin{equation}
\label{eq:Damping_function}
\Delta_\alpha(t):=\e^{-\Gamma_\alpha t}\theta(t)
\enspace,
\end{equation}
where $\theta(t)$ is the Heaviside step function.
The latter is introduced in order to have the integrations from $-\infty$ to $\infty$.
Further, the BOA is applied and the lifetimes of the intermediate states are assumed to depend
on the corresponding electronic level only, i.e.\ $\Gamma_\alpha=\Gamma_a$.
Then the interaction representation is employed in order to obtain the time evolution for the dipole operators
with respect to the electronic ground state,
see \Eqs{\ref{eq:interaction_representation},\ref{eq:scattering_op_definition}}.
In order to formulate a practical recipe involving classical MD methods, the dynamical classical limit is performed for nuclear degrees of freedom following \Sec{sec:DCL}.
Finally, the resulting RIXS amplitude possesses the form of a multi-time integral over the classical canonical average
with respect to the electronic ground state Hamilton function
\begin{align}
\label{eq:RIXS_final}
\mathcal{R}&(\Omega,\omega)=\frac{1}{ 2\pi} \intop_{-\infty}^{\infty}\mathrm{d}t\,\mathrm{e}^{\mathrm{i}\Omega t} 
\intop_{-\infty}^{\infty}\mathrm{d}\tau_{1}\,\mathrm{e}^{-\mathrm{i}\omega(t+\tau_{1})} 
\intop_{-\infty}^{\infty}\mathrm{d}\tau_{2}\,\mathrm{e}^{\mathrm{i} \omega\tau_{2}} 
   \nonumber \\
& \times \left  \langle\mathcal{W}_g(0)    {M}^{g \zeta}_{j}(t,0)\mathfrak{e}_{\zeta}
 \Delta_{j}(\tau_1) \mathfrak{u}^{\nu} {M}^{j}_{\nu f}(t+\tau_1,0)  \right.  \nonumber \\
& \left.  \times {M}^{f \eta}_{i}(\tau_2,0) \mathfrak{u}_{\eta}   \Delta_{i}(\tau_2)\mathfrak{e}^{\xi} {M}_{\xi g}^{i}(0,0)   \right \rangle
\enspace.
\end{align}
This expression for the RIXS amplitude together with the one for the XAS amplitude, \Eq{eq:XAS_final},  is the central theoretical result of this work.
These expressions coincide with the ones derived via optical response functions in Ref.~\cite{Karsten-arXiv-2016}.
Similar to the corresponding analogues in the UV/Vis domain~\cite{Heller1978, Heller-JCP-1979,Mukamel-Book} these expressions provide a general and unifying framework for simulating XAS and RIXS amplitudes.

\section{Methodology}
\label{sec:Methodology}
%
\subsection{Sampling and time-domain approaches}
\label{sec:sampling_approach}
As it has been stated in the introduction, the conventional approach to the X-ray amplitudes, \Eqs{\ref{eq:XAS},\ref{eq:Kramers_Heisenberg}}, is based on static single point calculations for structures sampled from statistical ensembles.
Note that the sampling amplitudes, which correspond to the limit of fixed nuclei of the respective time-domain expressions,
can be brought to the same functional form as the latter, see
Supplement.
This makes the analysis of the contributions of the nuclear dynamics, performed via comparison of the time-domain results to their sampling counterparts, convenient,  see \Sec{sec:RIXS}.
In order to conduct a fair comparison, the sampling procedure has been performed using identical large data sets as have been employed for the time-domain method.

\subsection{Tracing the states}
\label{sec:order_phase_problem}
In this section we aim at formulating practical recipes for evaluating the time-domain expressions for the
XAS and RIXS amplitudes, \Eqs{\ref{eq:XAS_final},\ref{eq:RIXS_final}}.
One notices that both are fully determined by the time evolution of the transition dipole moments, $D_{\xi b}^{a}(t)$, and the electronic energies, ${E}_{a}(t)$.
To reiterate, the evolution is carried out with respect to the electronic ground state potential energy surface.
%

Unfortunately, the functional dependence of the aforementioned ingredients on nuclear coordinates is not available for any realistic many-particle system.
A possible solution is to solve the TISE at each MD timestep independently using any established quantum-chemistry method,
which, however, would lead to two problems.
First, the numbering of the eigenvalues is arbitrary at any time instance.
Second, the phases of the eigenstates are ambiguous.
Ignoring these obstacles would ultimately destroy the dynamical correlation effects that are the essence of the developed formalism.
Thus, evaluating the time correlation functions requires tracing the states
along MD trajectories, which would yield a continuous evolution of the respective energies and phases.
Technically, the order and phases can be fixed at $t=0$ and then followed by identifying the states of the same character at future times.

In order to find a mapping of an unordered set of states $\{ \mKet{\tilde{b}(t)} \}$ onto the desired ordered set $\{ \mKet{a(t)} \}$, one can formally make the expansion 
\begin{equation}
  \mKet{a(t)}=\sum_{\tilde{b}}\langle\tilde{b}(t) \mKet{a(t)}\, \mKet{\tilde{b}(t)}
\enspace,
\end{equation}
assuming that both sets are complete.
Then, one can define a generalized permutation matrix, $\mMat{Y}(t)$, with elements $Y^{\tilde{b}}_a(t):=\langle\tilde{b}(t) \mKet{a(t)}$ that brings any unordered observable, $\tilde{A}_{\tilde{c}}^{\tilde{d}}(t)$,
to the correct order via ${A}_{a}^{b}(t)=Y^{b}_{\tilde{d}}(t)\tilde{A}_{\tilde{c}}^{\tilde{d}}(t)Y^{\tilde{c}}_a(t)$.
Note that matrix $\mMat{Y}(t)$ contains only one non-zero (complex unity) element per column {\it and} row by construction.

Since the correctly ordered set of states $\{ \mKet{a(t)} \}$ is not available, one can start at $t=0$ and to approximate the respective matrix elements as
\begin{equation}
  Y^{\tilde{b}}_a(\Delta t)\approx O^{\tilde{b}}_a(\Delta t,0) := \langle\tilde{b}(\Delta t)\mKet{a(0)} 
\enspace, 
\end{equation}
where $\Delta t$ is the next time instance, e.g.\ next MD timestep.
Since $O^{\tilde{b}}_a(\Delta t,0)$ might contain many non-zero contributions per row/column, one finds a maximal one in each column of $\mMat{O}(\Delta t,0)$, normalizes this (complex) number to unity and set the other to zero.
In order to verify the uniqueness of the result, the same procedure is performed row-wise.
If this self-consistency check is failed, it implies that the MD timestep is too large and has to be reduced.
The restored set of states at $t=\Delta t$ serves as a basis for ordering the states at $t=2\Delta t$ and so on
along the MD trajectory.
Since similar problems arise in surface hopping methods~\cite{Krylov1996}, the developed procedure might be of use there as well.

\subsection{An efficient evaluation of XAS and RIXS amplitudes}
The XAS amplitude, \Eq{eq:XAS_final}, is the Fourier transform of an equilibrium time correlation function.
Moreover, the RIXS amplitude, \Eq{eq:RIXS_final}, contains a multi-time correlation function and three time integrations, only one of those being decoupled from the other two.
This implies calculating a two-dimensional time integral on top of evaluating the correlation functions 
(each being
a nested time integration as well).
A handy way of making this numerically efficient can be adopted from vibrational spectroscopy~\cite{Ivanov-PCCP-2013},
where the stationarity of the canonical density is utilized implying that any time instance can be considered as the starting one.
Integrating over all initial times with the help of the convolution theorem leads to XAS and RIXS amplitudes in a form of products of the Fourier-transformed (indicated by the reversed hat) dressed dipole moments
\begin{equation}
\label{eq:TDM_FT}
 \FT{M}^{f}_g(\Omega) = \intop_{-\infty}^{\infty}\diff t\,\e^{-\i\Omega t}  {D}^{f}_{g}(t) \e^{\i \intop_{0}^{t} \diff \tau \Delta E_{fg}(\tau)}
  \enspace ,
\end{equation}
see Supplement for details.
%
%
Since the complex exponential depending on the energy gap oscillates on the electronic timescale, which is not accessible by MD, the common recipe for separating the highly oscillating contribution is employed~\cite{Kuehn-Book}.
One defines the mean transition frequency between two electronic states $f$ and $g$ as the average along the trajectory of a total length $T$, $\bar{\omega}_{fg}:=1/T\intop_{0}^{T} \diff \tau\Delta E_{fg}(\tau)$,
and the gap fluctuation as
$U_{fg}(\tau):=\Delta E_{fg}(\tau)-\bar{\omega}_{fg}$.
Inserting them into \Eq{eq:TDM_FT} leads to
\begin{equation}
\label{eq:frequency_shift}
 \FT{M}^{f}_g(\Omega) \equiv  \FT{\bar{M}}^{f}_g(\Omega-\bar{\omega}_{fg})
  \enspace ,
\end{equation}
where $\bar{M}^{f}_g(t,0):=D^{f}_{g}(t) \exp[\i \intop_{0}^{t} \diff \tau U_{fg}(\tau)]$.
The bars above the quantities generally indicate that the mean electronic gap has been removed and they evolve on the nuclear time scale and, thus, can be Fourier-transformed using the data provided by MD.

Finally, with the help of the convolution theorem the XAS amplitude can be expressed as a product
\begin{equation}
\label{eq:XAS_final_Fourier}
\mathcal{X}(\Omega)= \frac{1}{2\pi T} \mAve{ \FT{\bar{M}}^{g}_f(\bar{\omega}_{fg}-\Omega) \FT{\bar{\mathcal{M}}}^{f}_g(\Omega-\bar{\omega}_{fg})}
  \enspace ,
\end{equation}
where $\bar{\mathcal{M}}^{f}_g(t,0):=\mathcal{W}_g(t)D^{f}_{g}(t) \exp[\i \intop_{0}^{t} \diff \tau U_{fg}(\tau)]$;
note the different signs in the argument of the two functions in \Eq{eq:XAS_final_Fourier}.

Following the same approach, the RIXS amplitude can be reformulated as
\begin{widetext}
\begin{align}
\label{eq:RIXS_final_Fourier}
\mathcal{R}(\Omega,\omega)=\frac{1}{2\pi T} 
& \left \langle \intop_{-\infty}^{\infty} \right. \diff \omega_1  \FT{\bar{M}}^{g}_j\left(\bar{\omega}_{jg}-[\Omega-\omega_1 ]\right) \FT{\Delta}_j(\omega_1 ) \FT{\bar{M}}^{j}_f\left([\omega -\omega_1] - \bar{\omega}_{jf}\right)  \nonumber \\
& \times \left. \intop_{-\infty}^{\infty} \diff \omega_2   
\FT{\bar{\mathcal{M}}}^{i}_g\left([\Omega -\omega_2] - \bar{\omega}_{ig}\right)
\FT{\Delta}_i(-\omega_2 )  
\FT{\bar{M}}^{f}_i\left(\bar{\omega}_{if}-[\omega  - \omega_2] \right)
 \right \rangle
  \enspace ,
\end{align}
\end{widetext}
where $\FT{\Delta}_j(\omega):=1/(2\pi)\sqrt{\Gamma_j/\pi}(\Gamma_j+\i \omega)^{-1}$.
Here, one still has to perform a convolution due to the finite lifetime of the intermediate states.

To sum up, the calculation of both spectra is accomplished by performing single-variable integrations using the MD timestep.
It is worth mentioning that this structure of the final result naturally enables identifying the contribution from any particular initial, intermediate and/or final state.
Note also that identifying the contributions from the underlying geometrical motifs can be performed analogously to vibrational spectroscopy~\cite{Mathias-JCTC-2011,Mathias-JCTC-2012}.

%
\begin{figure}[th!]
  \includegraphics{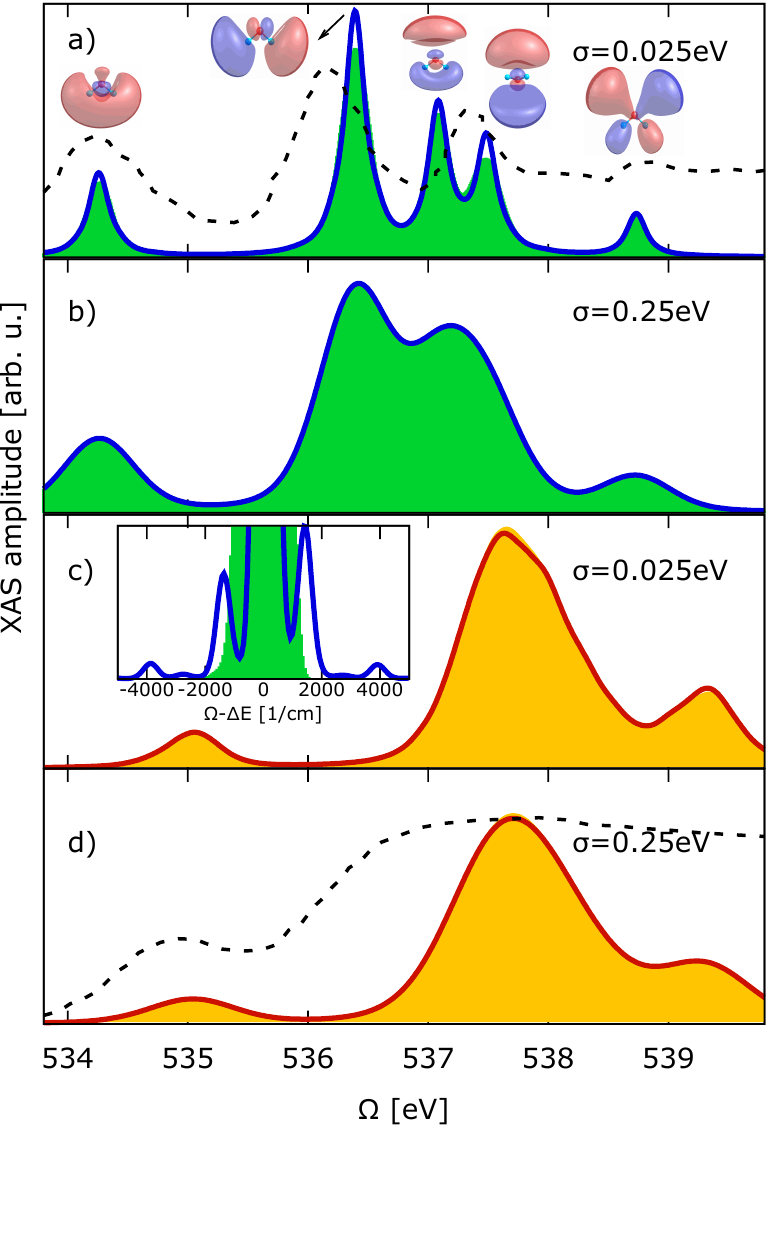}
\caption{\label{fig:XAS}
XAS amplitudes of water for a small ($\sigma=0.025$\,eV) and a large ($\sigma=0.25$\,eV) pulse width are shown in panels a), c) and b), d), respectively.
The gas phase spectra are presented in panels a) and b), whereas panels c) and d) are dedicated to that of bulk water.
Dashed lines represent normalized experimental data for bulk and gas phase water from Ref.~\citenum{Lange2013}.
Blue and red lines depict the time-domain approach results according to \Eq{eq:XAS_final_Fourier}, whereas filled (green and orange) curves correspond to the sampling method.
The unoccupied molecular orbitals to which the transition is performed are shown near the respective spectral peaks.
Inset zooms on the $1\mathrm{s}\rightarrow \sigma^*(2\mathrm{s})$ transition in the gas phase with the imposed infinite lifetime ($\Gamma_f \to 0$), where the frequency axis is shifted by the peak position $\Delta E =534.3$\,eV.
}
\end{figure}

\section{Computational details}
\label{sec:computational_details}
%
The MD simulations have been performed using the \textsc{Gromacs} program package ver.~4.6.5~\cite{GROMACS} employing the anharmonic qSPC/Fw water model with a Morse potential for the O-H intra-molecular potential~\cite{Paesani-JCP-2006}.
The ``standard protocol''~\cite{Ivanov-PCCP-2013} for calculating spectra in the canonical ensemble has been used, that is, a set of uncorrelated initial conditions has been generated from an $NVT$ MD run with the target temperature of $300$\,K imposed by the Langevin thermostat.
These initial conditions have been further used as starting points for simulating microcanonical ($NVE$) trajectories, each $0.5$\,ps long, yielding a spectral resolution of $\approx 8$\,meV.
%
The MD timestep of $0.1$\,fs has been used to provide a 
successful tracing of the states as described in \Sec{sec:order_phase_problem}.
Note that for a present system we could use a timestep of $0.5$\,fs, but we opted to take the smallest employed timestep for the sake of numerical accuracy.
The spectra calculated along the $NVE$ trajectories have been averaged to yield the result for the  canonical ensemble.

Two systems have been considered: a) an isolated water molecule, referred to as gas phase water; b) a cubic box consisting of 466 water molecules under periodic boundary conditions with the box edge length of $2.4\,$nm, which corresponds to a density of $1.0\,$g/cm$^{3}$, referred to as bulk water.

The TISE (\Eq{eq:BOA2}) has been solved for each MD snapshot at the level of the ground state density functional theory with the Perdew-Burke-Ernzerhof functional~\cite{PBE_1996} using the ORCA ver.~3.0.3 program package.~\cite{orca_2012} 
Tight SCF convergence criteria ($10^{-7}$~Hartree) and a standard grid (ORCA grid3) have been employed.  
The def2-QZVPP basis set for oxygen and hydrogen~\cite{def2_2005} together with (5s5p)/[1s1p] generally contracted Rydberg functions on oxygen have been used.
Rydberg contractions have been obtained as atomic natural orbitals~\cite{ano_1987,roos_1996} constructed of primitives with universal exponents (see Ref.~\citenum{Kaufmann_1989}).
Such a small Rydberg basis does not allow one to reproduce the high-energy tail of the absorption spectrum~\cite{naeslund_JPCA_2003} but enables description of the lowest states just above the core-excitation threshold.
The energies of the singly-excited valence and core states have been approximated by the differences of the respective Kohn-Sham orbital energies; the corresponding dipole transition moments have been calculated with respect to these orbitals.~\cite{lee_JACS_2010}
This quasi--one-electron method is known to be a reasonable compromise between accuracy and efficiency, as has been demonstrated for various kinds of systems.~\cite{Hennies2005,naeslund_JPCA_2003,lee_JACS_2010,Lassalle-Kaiser2013,Pollock2013}

In order to achieve a feasible treatment of the bulk system, a hybrid quantum mechanics/molecular mechanics (QM/MM) partitioning has been performed.
For each snapshot {\it and} each selected molecule individually, the whole simulation box has been translated to the mass centre frame of that molecule which has been further treated via the QM method.
The atoms of all surrounding molecules have been considered as classical point charges that enter the QM calculation via a static external potential.

Since the QM calculations are to be performed for an isolated (hence non-periodic) system,  surface effects due to the finite box have been ``smoothed out" by applying a spherically symmetric cutoff function $f(r)$ to the charges, i.e.
\begin{equation}
f(r):=\begin{cases}
1, r<R_{\mathrm{in}}\\
\frac{2(r-R_{\mathrm{in}})^3}{(R_{\mathrm{out}}-R_{\mathrm{in}})^3}-\frac{3(r-R_{\mathrm{in}})^2}{(R_{\mathrm{out}}-R_{\mathrm{in}})^2}+1, R_{\mathrm{in}}\leq r \leq R_{\mathrm{out}}\\
0, r > R_{\mathrm{out}} \\
\end{cases}
\enspace ,
\end{equation}
with a smoothing region $[R_{\mathrm{in}},R_{\mathrm{out}}]$, where $r$ is the distance from the origin to the mass center of a surrounding water molecule.
We have employed the values $R_{\mathrm{in}}=5$\,\AA\ and $R_{\mathrm{out}}=8$\,\AA \ to safely include all charges in the first solvation shell.
In order to preserve the  charge neutrality of the entire system the same function value has been applied 
to all point charges belonging to the same molecule.
%
%
Convergence has been reached for considering the trajectories of one 120 molecules for the gas phase water and 70 for the bulk water.

In order to mimic the finite width $\sigma$ of the exciting light pulse, as given by experimental conditions, the spectra, \Eqs{\ref{eq:XAS_final},\ref{eq:RIXS_final}},
have been convoluted with normalized Gaussian functions $\exp[-\Omega^2/(2\sigma^2)]/(\sqrt{2\pi\sigma^2})$ in $\Omega$-direction
with $\sigma=0.025$\,eV and $\sigma=0.25$\,eV as typical small and large widths, respectively.~\cite{Milne2014}
Additionally, the XAS amplitude has been convoluted with normalized Lorentzian functions $\Gamma_f/\pi \cdot [\Gamma_f^2+\Omega^2]^{-1}$ taking into account the finite lifetime of the core-excited final states.
For the results in \Sec{sec:results}, the value $\Gamma_a=0.25$\,fs$^{-1}$, as a typical decay rate for core holes \cite{Hjelte2001}, has been used for all final states entering the XAS amplitude as well as for all intermediate states entering the RIXS amplitude, where the finite lifetime is already accounted for by construction, see \Eqs{\ref{eq:Kramers_Heisenberg},\ref{eq:RIXS_final}}.

The number of considered states that contribute to the spectra is determined by the absorption and emission frequency range of interest.
Here, $\Omega\in[532.8,540.8]$\,eV and $\omega\in[509.8,540.8]$\,eV requires the consideration of 31 states for both water setups.
Note that all spectra have been shifted {\it globally} by 24.8\,eV such that the peak structure roughly matches the experimental data for bulk and gas phase water~\cite{Lange2013}.
Finally, the data shown in \Sec{sec:results} corresponds to an average over orthogonal polarizations of the incoming and the emitted light~\cite{monson1970,Luo-PRA-1996} owing to the isotropy of the gas and liquid phases.
%

\section{Results and discussion}
\label{sec:results}
The proposed methodology has been first suggested by us in Ref.~\citenum{Karsten-arXiv-2016} and applied to K-edge oxygen spectra of a gas phase water molecule.
Nuclear correlation effects have been demonstrated to be essential for second-order X-ray spectroscopy.
In particular, RIXS has turned out to be a sensitive technique for the effects in question,
whereas XAS exhibited almost no traces of the underlying nuclear dynamics.
However, the origin of the observed phenomena has not been analyzed and will be the main concern here.

\subsection{XAS}
\label{sec:XAS}
In this section, the results for the XAS amplitude obtained via various simulation scenarios are discussed and compared against each other.
We would like to stress again that the focus is put on the differences due to nuclear dynamics rather than on the peculiarities of the water spectra themselves, see Ref.~\citenum{Fransson-CR-2016} and references therein.

In \Fig{fig:XAS} the XAS amplitudes for the small ($\sigma=0.025$\,eV) and the large ($\sigma=0.25$\,eV) width of the excitation pulse provided by the sampling and the time-domain approach are shown for   gas phase as well as for   bulk water. 
Before comparing the spectra, we assign the peaks in the XAS to the underlying transitions in order to connect to the energy level structure of the water molecule.
The respective unoccupied molecular orbitals
are exemplified near the spectral peaks.
In particular, the first two peaks (534.3\,eV and 536.4\,eV) correspond to the $1\mathrm{s}\rightarrow \sigma^*(2\mathrm{s})$ and $1\mathrm{s}\rightarrow \sigma^*(2\mathrm{p})$ transitions, respectively.
Other peaks can be attributed to the transitions from 1s to 3p Rydberg orbitals of oxygen.

Comparing the results of the sampling and correlation approaches for the gas phase water molecule with the small linewidth, \Fig{fig:XAS}a), one sees that there are subtle but evident differences in intensities for all the peaks apart from the one with the highest energy.
Their origin can be clearly traced back by setting the lifetime of the final core-excited states to infinity ($\Gamma_f \to 0$),
see inset.
Here, observed pronounced side bands can be directly related to the vibrational modes of the water molecule,
in particular to the bending and stretching ones, which have the frequencies of $\approx\! 1500$\,\cm\ and 
$\approx\! 3800$\,\cm, respectively.
We stress that these side bands can not be provided by the sampling approach due to its intrinsic limitations.
Nonetheless, all these discrepancies disappear at the large pulse width, see \Fig{fig:XAS}b).
Note that the two spectra have the same area by construction, see Supplement for a proof, and thus increasing the width of the convoluted Gaussian naturally eliminates the differences between the two approaches.
Therefore, one can view this coincidence as the cross-check for the implementation.

The same comparison is performed for bulk water, see panels c) and d) in \Fig{fig:XAS}.
Here, the differences between the pulse widths and especially between the methods are vanishing.
Interestingly, the discrepancies are negligibly small even for infinite lifetimes, see Fig.~S1 in the Supplement.
This illustrates the statement that XAS is not a very sensitive observable for nuclear correlation effects.

To sum up, nuclear correlations do not influence XAS amplitudes for bulk water under any circumstances, whereas for gas phase the (small) differences are seen only if the pulse widths and/or lifetime broadening of the final states are particularly small.
In principle, the RIXS amplitude contains more information and, thus, could be more promising for observing nuclear effects, as will be shown in the next section.

\begin{figure}[tb]
	\begin{center}
		\includegraphics{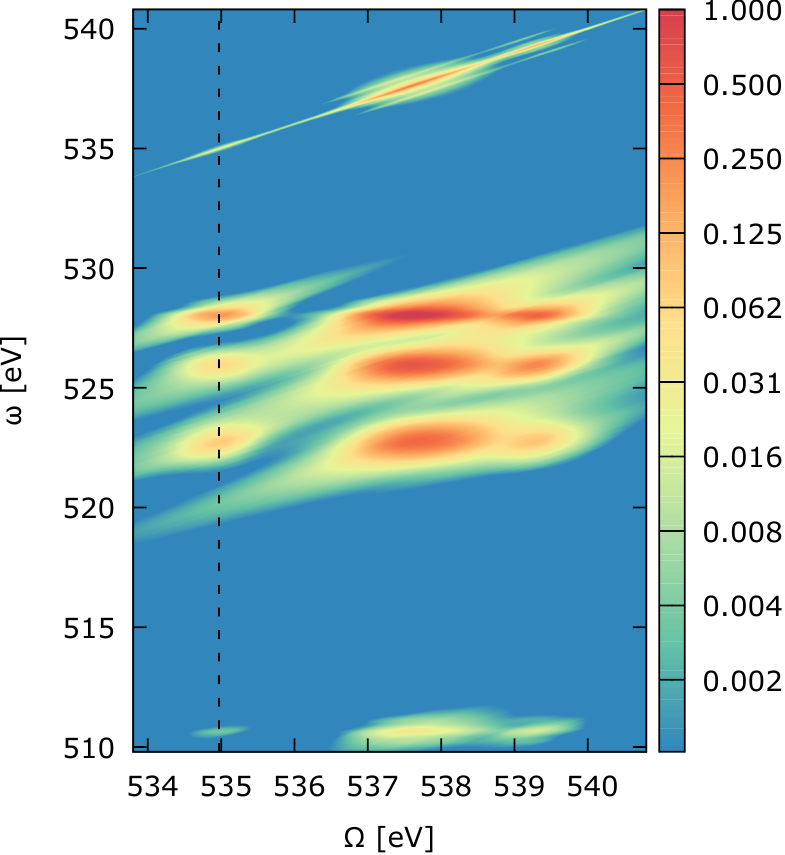}
		\caption{\label{fig:RIXS_2d}
			2D RIXS spectrum for bulk water (small pulse width) obtained by means of the time-correlation approach, \Eq{eq:RIXS_final_Fourier}.
			%
			The dashed vertical line indicates the position of the cut depicted in \Fig{fig:RIXS_cut}.
		}
	\end{center}
\end{figure}

\subsection{RIXS}
\label{sec:RIXS}

\begin{figure}[h!]
 \begin{center}
  \includegraphics{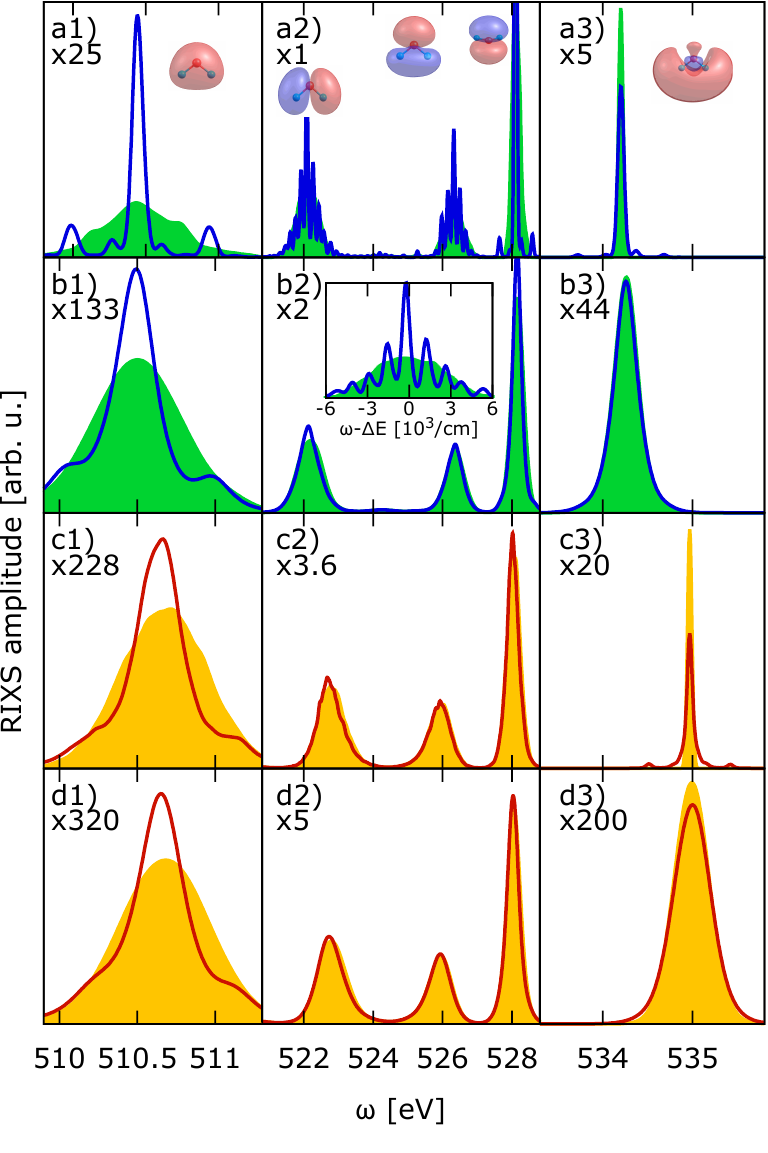}
\caption{\label{fig:RIXS_cut}
Cuts through RIXS spectra for various simulation scenarios.
The excitations frequencies are fixed at 534.2\,eV and 535.0\,eV for the gas phase and the bulk, respectively.
The colour-code and the panel structure are the same as in \Fig{fig:XAS}.
Panels a) and c) correspond to $\sigma=0.025$\,eV whereas b) and d) to $\sigma=0.25$\,eV. 
Each panel is split into three sub-panels according to the spectral ranges that exhibit intensity (note multiplicative factors therein), see \Fig{fig:RIXS_2d}. 
Inset zooms on the left peak in panel~a2) with $\Delta E=522.1$\, eV.
}
 \end{center}
\end{figure}

\begin{figure}[tbh]
 \begin{center}
  \includegraphics{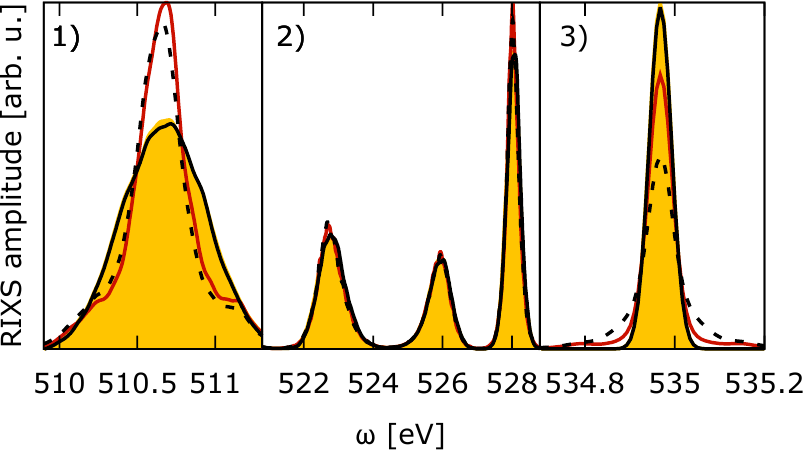}
 \end{center}
 \caption{\label{fig:compare_approx}The accuracy of the Condon approximation. The same cut as in \Fig{fig:RIXS_cut} panels c) is shown with the same colour code corresponding to bulk water with $\sigma=0.025$\,eV. Black dashed and solid curves correspond to the reference results without Condon approximation for the time-domain approach and the sampling, respectively.  }
\end{figure}

In \Fig{fig:RIXS_2d} a 2D spectrum for bulk water obtained according to \Eq{eq:RIXS_final_Fourier} is shown.
Although it gives an overall impression about the spectral shape in the entire excitation and emission ranges, it is hard to make a quantitative analysis on its basis.
Therefore, we consider a particular cut for a fixed excitation frequency $\Omega$ that corresponds to a maximum of the spectral peak assigned to the $1\mathrm{s}\rightarrow \sigma^*(2\mathrm{s})$ absorption transition, see vertical line in \Fig{fig:RIXS_2d};
note that the particular excitation frequencies are different for the gas phase (534.2\,eV) and bulk (535.0\,eV) cases.
Note further that the tendencies observed for cuts at different excitation frequencies are qualitatively similar, see Ref.~\citenum{Karsten-arXiv-2016} and Supplement.

Three spectral ranges corresponding to peaks in the aforementioned cut are shown for various simulation scenarios in \Fig{fig:RIXS_cut}.
These peaks can be related to transitions from the intermediate (core-excited) state to final (ground or valence-excited)  states, see the respective orbitals from which the emission takes place in panels a).
In particular, the peak in panel~a1) can be assigned to the $\sigma(2\mathrm{s}) \rightarrow 1\mathrm{s}$ core-hole refill.
In panel~a2), the peaks at 522.0, 526.4, and 528.1\,eV stand for the refill from bonding $\sigma(2\mathrm{p})$ and two lone-pairs n(2p), respectively.
Panel~a3) contains the elastic peak, that is, the refill from the antibonding $\sigma^*(2\mathrm{s})$ orbital populated in the first step of the RIXS process.
In panels a1)-a3) one sees pronounced differences between the methods in case of the small pulse width.
A prominent vibronic structure is observed for all peaks with the frequencies easily attributed to vibrational normal modes, see, e.g., inset where the fingerprints of the bending mode can be clearly seen.
Although these structures  disappear for the large pulse width, see panels b) therein, differences in intensity remain for elastic and for the lone-pair peak with the highest energy (528.1~eV).
Moreover,  the energetically lowest transition in panel~b1) still exhibits a pronounced peak structure when computed with the time-correlation approach.
Importantly, the results for bulk water still reveal noticeable differences between the methods for all pulse widths considered, see panels c1), c3), d1) and d3).
In particular, the sampling approach overestimates the intensity of the elastic peak and underestimates the intensity of the inelastic ones and, in addition, reveals no vibronic progressions.
This underlines the statement that RIXS spectra are sensitive to correlation effects in the nuclear dynamics.
Most importantly, one sees clear traces of nuclear dynamics at all realistic experimental conditions considered.

In order to shed light on the origin of the observed deviations, 
we consider all possible sources, see \Eq{eq:RIXS_final}, that is, the time correlation of the transition dipoles and that of the energy gaps.
The former corresponds to the effects beyond the Condon approximation, which can be easily elucidated by setting the dipoles to their values at, e.g., $t=0$.
It turns out that the deviations due to the Condon approximation are small for inelastic peaks, see \Fig{fig:compare_approx}, and thus cannot be responsible for the substantial differences observed in RIXS spectra.
This implies that these differences are caused by the time dependence of the energies involved.
Interestingly, the aforementioned deviations for the elastic peak are notable only for correlated spectra computed here.
Importantly, the Condon approximation increases the intensity of the elastic peak, whereas correlation effects suppress it.
Since the two act in opposite directions, one can employ the Condon approximation results for the analysis,
as the true non-Condon differences between the sampling and correlation methods can be only more pronounced.

To investigate the role of the energy time-dependencies, we consider a particular RIXS channel,
$g \to i \to f$, and impose the Condon approximation as it has been justified for inelastic peaks for the present system.
The corresponding RIXS amplitudes for the sampling, $\mathcal{R}^S$, and the time-correlation approach, $\mathcal{R}^C$,
taken at the mean transition frequencies $\bar{\omega}_{ig}$ and $\bar{\omega}_{if}$ can be reduced to 
\begin{align}
\label{eq:R_reduced}
\mathcal{R}^\mathrm{S/C}&(\bar{\omega}_{ig},\bar{\omega}_{if})\propto 
\intop_{0}^{\infty} \diff t  \, C_{gif}^{\mathrm{S/C}}(t)
\enspace,
\end{align}
where
\begin{subequations}
\begin{equation}
\label{eq:C_S}
C_{gif}^{\mathrm{S}}(t) := \e^{-\sigma^2 t^2 /2}  \Re \mAve{\e^{\i U_{gf}(0)t} | \chi_{if}^{\mathrm{S}}(0) |^2 }
\end{equation}
\begin{equation}
\label{eq:C_C}
C_{gif}^{\mathrm{C}}(t):= \e^{-\sigma^2 t^2 /2}  \Re \mAve{\e^{\i \intop_{0}^{t} \diff \tau U_{gf}(\tau)}  \chi^{\mathrm{C}*}_{if}(t) \chi^\mathrm{C}_{if}(0) }
\enspace ,
\end{equation}
\end{subequations}
as it is shown in the Supplement. 
The functions $\mathcal{R}^\mathrm{S/C}(\bar{\omega}_{ig},\bar{\omega}_{if})$ approximately describe the heights of the corresponding spectral peaks at maximum.
Here,
\begin{align}
\label{eq:chi_if}
\chi_{if}^{\mathrm{S/C}}(t):=\intop_{-\infty}^{\infty} \diff \tau \, L_{if}^\mathrm{S/C}(\tau ;t)
\enspace ,
\end{align}
with 
\begin{subequations}
\begin{equation}
\label{eq:L_S}
L_{if}^\mathrm{S}(\tau ; 0):=\exp[ \i U_{if}(0)\tau] \Delta_i(\tau)
\end{equation}
\begin{equation}
\label{eq:L_C}
L_{if}^\mathrm{C}(\tau ; t):=\exp[ \i \intop_{t}^{t+\tau} \diff \bar{\tau} \,U_{if}(\bar{\tau})] \Delta_i(\tau)
\enspace,
\end{equation}
\end{subequations}
where $\Delta_i$ is the damping function defined in \Eq{eq:Damping_function}, and $U_{ab}(t)$ are the gap fluctuations;
note that $\chi_{if}^{\mathrm{S}}(t)$ is evaluated only at $t=0$.
Importantly, $C_{gif}^{\mathrm{S}}(t)$ consists of the averaged product of the phase w.r.t.\ $U_{gf}(0)$ and the squared value of $\chi_{if}^\mathrm{S}(0)$, which is just a number.
In contrast, $C_{gif}^{\mathrm{C}}(t)$ is constructed from the averaged product of the two expressions, both involving autocorrelation functions, since the phase factor in \Eq{eq:C_C} can be treated as an autocorrelation function in the framework of the cumulant expansion~\cite{Mukamel-Book,Kuehn-Book}.
Note that the two autocorrelation functions involve \textit{different} gap fluctuations, i.e.\ $U_{if}$ and $U_{gf}$.

\begin{figure}[h!]
  \includegraphics{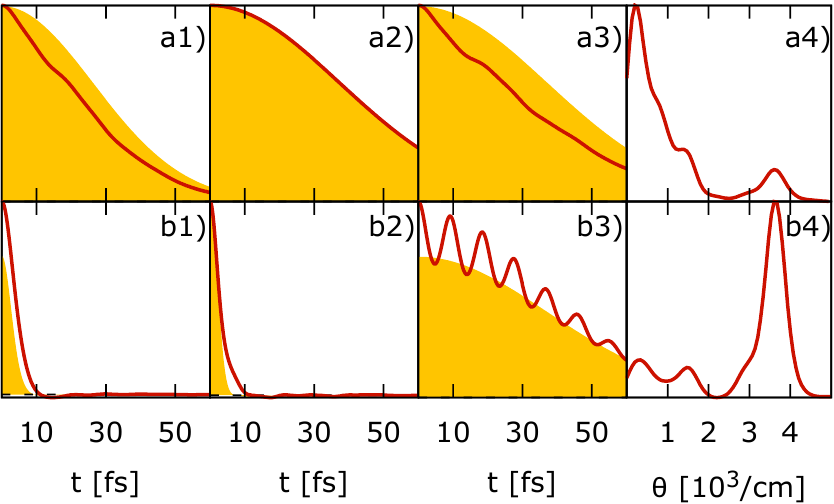}
 \caption{\label{fig:correlation_functions}
The origin of different peak intensities. The colour code can be taken from \Fig{fig:RIXS_cut}. Row a) corresponds to the intensity of the elastic peaks in Condon approximation depicted by the orange and red curves in  panel 3), \Fig{fig:compare_approx}. The lower row b) is related to the $\sigma(2\mathrm{s}) \rightarrow 1\mathrm{s}$ transition, see \Fig{fig:compare_approx} panel 1). The first column contains the time-dependent functions that determine the peak intensities, see \Eqs{\ref{eq:C_S},\ref{eq:C_C}}. The second and third columns feature the first and the second part of the factorization in \Eq{eq:factorisation}, respectively, as well as the sampling counterparts. In the fourth column one can see the spectral densities (averaged absolute value of the Fourier transformed gap fluctuations), where the results are convoluted with a normalized Gaussian ($\sigma=0.025$~eV).}
\end{figure}

To proceed, the following approximate factorization is performed
\begin{align}
\label{eq:factorisation}
C_{gif}^\mathrm{C}(t) & \approx \Re \left[ \e^{-\sigma^2 t^2 /4}  \mAve{\e^{\i \intop_{0}^{t} \diff \tau U_{gf}(\tau) }} \right] \nonumber \\
&\times \Re  \left[ \e^{-\sigma^2 t^2 /4}  \mAve{\chi^{\mathrm{C}*}_{if}(t) \chi_{if}^\mathrm{C}(0) } \right]
\enspace;
\end{align}
here the Gaussian has also been split into two in order to smooth both parts in a similar way. 
The same factorization is then performed for $C^\mathrm{S}_{gif}(t)$ defined in \Eq{eq:C_S}.
These factorizations are possible due to the absence of correlations between the two parts as can be shown numerically and seen from \Fig{fig:correlation_functions}.
Thus, the aforementioned averaged product of the two factors has now become the product of the averages.

The relevant quantities are summarized in \Fig{fig:correlation_functions}, whose two rows correspond to the peaks exhibiting the most pronounced differences in the RIXS spectra, that is, the elastic one and the peak with the lowest emission energy, see \Fig{fig:RIXS_cut}; the analysis for the other three peaks is shown in the Supplement.
The first column contains $C^\mathrm{S/C}_{gif}$, the second and third columns depict the first and second parts in \Eq{eq:factorisation}, respectively, as well as the corresponding sampling counterparts.
The fourth column shows the averaged absolute values of the Fourier-transformed gap fluctuations, i.e., the spectral density function.
The intensity differences in question are given by the difference of the \textit{areas} under the $C^\mathrm{S/C}$ curves in the first column therein.
To reiterate, although for the elastic peak the Condon approximation is not justified, going beyond it will make the effect in question even stronger, as becomes apparent from \Fig{fig:compare_approx}.

For elastic scattering $g=f$, and hence the first part in \Eq{eq:factorisation} reduces to $\exp[{-\sigma^2 t^2 /4}]$ by construction, as $U_{gg}(\tau)\equiv 0$; the same naturally applies to its sampling counterpart, see panel~a2).
Therefore, $C^\mathrm{S}$ (shaded area in panel~a1)) exhibits a Gaussian decay stemming from the excitation pulse.
The solid curve therein, $C^\mathrm{C}$, contains in addition {\it correlation} effects due to the second part of \Eq{eq:factorisation} shown in panel~a3), that shrink the area under the curve~a1),
and are thus responsible for the observed intensity difference in the spectra, \Fig{fig:compare_approx} panel 3).

The same analysis performed for the inelastic peak immediately suggests that the shape of $C^\mathrm{S/C}$, panel~b1), is dictated almost exclusively by the first part in \Eq{eq:factorisation} presented in panel~b2).
The observed rapid decay ($\approx\!\!10\,$fs) in turn implies that the peculiarities of the time-dependence of the correlation function constituting the second part, panel~b3), are irrelevant.
Thus, its value at $t=0$, which can be notably different for the two approaches, controls the magnitude of $C^\mathrm{S/C}$ and, hence, the signal intensity at maximum.

One can show on the basis of a simple harmonic model, that in the limit of fast nuclear vibrations the sampling method would underestimate $\chi_{if}^\mathrm{C}(0)$ and thus $C^\mathrm{S/C}(0)$, whereas the two methods coincide in the low-frequency limit, see Supplement.
This implies that \textit{inelastic} spectral intensities would be underestimated as well.
Practically one can expect this underestimation
to be strong if the gap fluctuation has more contributions from high-frequency modes than from low-frequency ones.
Indeed, the spectral density shown in panel~b4) strongly supports this scenario, see Supplement for the respective contributions for other peaks.
Furthermore, the similar values of $C^\mathrm{S/C}(0)$ for the elastic peak in panel~a1) directly follow from the fact that the corresponding
spectral density has its major contribution in the low frequency range, see panel~a4).
Nonetheless, for elastic peaks not only the value at zero is important but the entire shape of $C^\mathrm{S/C}(t)$, and thus the overall conclusion for elastic peaks can not be drawn on this basis.

Interestingly, the impact of the Condon approximation on spectra can be explained by the above analysis.
In particular, the time-dependence of the transition dipoles, which is ignored by the approximation, serves as an additional source for dephasing, decreasing the elastic peak intensity even further, see \Fig{fig:compare_approx}.
Since for inelastic peaks just the value of $C^\mathrm{S/C}(0)$ matters, the faster dephasing is expected to be almost irrelevant and thus, the Condon approximation has proven itself reliable in this case.

To resume, the differences in intensities between the two approaches for the system studied are determined by the second part of \Eq{eq:factorisation}, which is a clear trace of nuclear dynamics.

\subsection{Limitations of the method}
\label{sec:limitations}

Let us recapitulate the approximations that have been employed on the way from the coupled nuclear-electronic Schr\"odinger equation to the final result, \Eqs{\ref{eq:XAS_final},\ref{eq:RIXS_final}}.
%
The first one is the BOA, which leads to ignoring non-adiabatic effects that can be important in a typical scenario when the density of core-excited states is substantial.
In general, having the continuous time evolution of the electronic wavefunction opens the doorway for taking these effects into account, and thus this is not a principal deficiency.

The second issue is the electronic structure method employed.
%
Here, we use ground state Kohn-Sham orbitals for the excited states and thus inevitably neglect electronic relaxation,
differential correlation, and possible multi-configurational nature of the wavefunction.
%
%
Importantly, the developed approach is independent of the electronic structure method and, thus, utilizing a truly correlated technique for the excited states~\cite{Helgaker-Book}, such as the multi-configurational self-consistent field approach, would mitigate the problem.~\cite{Josefsson-JCPL-2012,Bokarev-PRL-2013,Bokarev-JPCC-2015,Wernet-Nature-2015}

The last but not the least approximation is the classical one.
In particular, the classical treatment of the scattering operator, \Eq{eq:scattering_op_definition}, which does not have in principle the correct classical limit, is, in our opinion, the most severe approximation employed.
The replacement of the operator $\Delta \mOp{E}_{a0}$ by a number, leads to the complete loss of information about the dynamics in the excited state $\mKet{a}$, and leaves us with a simple phase factor.
The consequences of the dynamical classical limit have been extensively studied by Berne et al.~\cite{Egorov-JCP-1998,Rabani-JCP-1998}.
%
This approximation can cause wrong frequencies and shapes of the vibronic progressions in certain physical situations, whereas the envelopes of the vibronic bands are reproduced reasonably well.
%
%
In addition, the phenomenological dephasing model used, \Eq{eq:Damping_function}, leaves cases that exhibit intricate large-amplitude dynamics in the excited state
outside reach. 
%
Note that the classical approximation for the dipoles appears to be not so important, since the Condon approximation holds reasonably well.
We believe that the observed symmetry of the vibrational progressions with respect to 0-0 vibronic transition is also due to the classical limit, which corresponds to a temperature much larger than a vibrational quantum.
This can be understood in terms of a simple Huang-Rhys model, where such a symmetry emerges as a result of equally populated nuclear levels of the initial electronic state.

To resume, the only unsurmountable approximation that belongs to the method itself is the classical limit and thus is the main target for future improvements.



\section{Conclusions}
\label{sec:conclusions}

In order to include correlation effects from the underlying nuclear dynamics into theoretical X-ray spectroscopy,
the time-domain approach to XAS and RIXS spectra has been rigorously developed and tested on gas phase and bulk water.
The derivation has been carried out here from the Schr\"odinger-picture expressions and in Ref.~\citenum{Karsten-arXiv-2016} 
from optical response functions. 
It has been shown that at realistic (experimental) conditions the impact of nuclear dynamical effects on XAS amplitudes of water is fairly small.
However, the RIXS spectra have turned out to exhibit pronounced signatures of the nuclear dynamics that have been traced down to the particular underlying effects in all cases studied.
The difference between the sampling and the approach presented here has turned out to be caused by nuclear correlation effects.
The observed intensity differences have been rationalized on the basis of a simple harmonic model and connected to the high- and low-frequency contributions to the spectral density.
The practical limitations of the method and  underlying approximations, with the dynamical classical limit being the most important one, have been analyzed and discussed.
%
%
Despite the deficiencies of the method, it represents a step forward over the conventional approaches treating the system in full complexity and provides a reasonable starting point for further improvements.

\begin{acknowledgments}
We acknowledge financial support by the Deanship of Scientific Research (DSR), King Abdulaziz University, Jeddah, (grant No.\ D-003-435) and the Deutsche Forschungsgemeinschaft (KU~952/10-1 (S.K.), IV~171/2-1 (S.D.I.)).
Special thanks go to Fabian Gottwald for technical assistance with the MD simulations of water.
\end{acknowledgments}


\bibliography{RIXS-add,WaterSolutions}
\end{document}